\documentclass[a4paper, 10pt, conference]{ieeeconf}      % Use this line for a4 paper

\IEEEoverridecommandlockouts                              % This command is only needed if 
\usepackage[pdftex]{graphicx,xcolor}% for pdflatex
\usepackage[pdftex]{graphics}
\usepackage[fleqn]{amsmath}
\usepackage{newtxtext}
\usepackage[varg]{newtxmath} 
\usepackage{ifpdf}
\usepackage{bm}
\usepackage{url}
 \usepackage{cases}
\usepackage{color}
\usepackage{multirow}
\usepackage[flushleft]{threeparttable}

\newcommand\NEXT{F}

\newcommand\SCOST{S}

\newcommand\PRIO[2]{\prec_{#1,#2}}

\title{\LARGE \bf
Decision Making for Autonomous Vehicles at Unsignalized\\ Intersection in Presence of Malicious Vehicles
}

\author{Sasinee Pruekprasert$^{1}$, Xiaoyi Zhang$^1$, J\'er\'emy Dubut$^{1,2}$, Chao Huang$^1$, Masako Kishida$^1$% <-this % stops a space
\thanks{The authors are supported by ERATO HASUO Metamathematics for Systems Design Project (No. JPMJER1603), JST; J. Dubut is also supported by Grant-in-aid No.~19K20215, JSPS.}% <-this % stops a space
\thanks{$^{1}$The authors are with the National Institute of Informatics, Hitotsubashi 2-1-2, Tokyo 101-8430, Japan\newline
        {\small \{sasinee, xiaoyi, dubut, chao\_huang, kishida\}@nii.ac.jp}}%
\thanks{$^{2}$J.D. is also affiliated with the Japanese-French Laboratory for Informatics}%
}

\begin{document}

{\onecolumn\large 
\noindent\textcopyright\ 2019 IEEE.  Personal use of this material is permitted.  Permission from IEEE must be obtained for all other uses, in any current or future media, including reprinting/republishing this material for advertising or promotional purposes, creating new collective works, for resale or redistribution to servers or lists, or reuse of any copyrighted component of this work in other works.}
\newpage
\twocolumn

\maketitle
\thispagestyle{empty}
\pagestyle{empty}
%\setlength{\textfloatsep}{7pt}

%%%%%%%%%%%%%%%%%%%%%%%%%%%%%%%%%%%%%%%%%%%%%%%%%%%%%%%%%%%%%%%%%%%%%%%%%%%%%%%%
\begin{abstract}

In this paper, we investigate the decision making of autonomous vehicles in an unsignalized intersection in presence of malicious vehicles, which are vehicles that do not respect the law by not using the proper rules of the right of way. Each vehicle computes its control input as a Nash equilibrium of a game determined by the priority order based on its own belief: each of non-malicious vehicle bases its order on the law, while a malicious one considers itself as having priority. To illustrate the effectiveness of our method, we provide numerical simulations, with different scenarios given by different cases of malicious vehicles.
\end{abstract}

%%%%%%%%%%%%%%%%%%%%%%%%%%%%%%%%%%%%%%%%%%%%%%%%%%%%%%%%%%%%%%%%%%%%%%%%%%%%%%%%
\section{Introduction}

Autonomous vehicles have a promising future of making transportation time effortless and enabling the driver to partake in other activities and therefore change everyday life across the world. The greatest strength of autonomous vehicle is to avoid traffic accidents caused by human error, and thus to improve safety on roadways \cite{bagloee2016autonomous}. With the development of autonomous vehicles, the focus in research moves towards complex traffic scenarios such as intersections. %{\color{red}with aggressive driving behaviour}. 

Intersections, at which multiple roads and opposing directions of traffic meet together, are designed to prevent traffic jams by easing the flow of traffic. 
The intersections can be divided into signalized intersections and unsignalized intersections \cite{timmurphy.org}. The signalized intersections have signs and signals to regulate traffic while the unsignalized intersections do not. The unsignalized intersections are typically found where the traffic flow is low. However, they usually have higher collision frequencies than the signalized intersections due to driver's indecision. One report of National Highway Traffic Safety Administration (NHTSA) in 2011 indicates that 40$\%$ of car collisions in the U.S. happen at intersections and 60$\%$ of them are related to unsignalized intersections \cite{hubmann2017decision}.  It is expected that the autonomous vehicle can efficiently avoid intersection mishap by making appropriate decisions and orchestrating proper actions without driver's intention.

The decision making by autonomous vehicle at unsignalized intersections is a critical problem. Several approaches have been recently proposed.  One common way is to investigate the human behaviour in unsignalized intersection in order to create a naturalistic decision process \cite{liu2007human, gadepally2014framework}.  For example,  \cite{de2017decision} proposes a human-like decision-making algorithm based on the human drivers' data. A cooperative approach is developed for collision avoidance at intersections based on formal control theoretic techniques \cite{hafner2013cooperative}. However, the approach requires vehicle-to-vehicle communication and thus is not scalable to general intersection handling. The principle of reachability-based decision-making approach, which predicts the probability of future collisions using error propagation, is applicable to frontal collision avoidance at intersections \cite{de2014collision}. However, this approach may lead to conservative results. An intention-aware decision-making model for unsignalized intersections is studied in \cite{song2016intention}, where
 drivers' intentions are modelled using two-layered hidden Markov models. 
 However, 
 %no uncertainty is considered during the planning as 
 the intention estimation is deterministic.

%Two layered hidden Markov models (HMMs) are used to model driving intention including motion intention (e.g., turn right) and interaction intention (e.g., predicted status of other vehicles), then the decision making model is designed to calculate the optimal policy.

%\begin{figure}[t!]
%	\begin{center}	{\includegraphics[width=0.3\textwidth]{Fig/Picture3}}
%		\caption{Unsignalized intersection}
%		\label{fig:bicycle}
%	\end{center}
%	%\vspace{-1cm}
%\end{figure}

This paper uses game theory for the decision making. Game theory is the study of decision making where the players (e.g. vehicles) make choices by maximising their own expected utility based on the possible actions of other players. Thus, game theory provides a promising framework for scenarios where interactions are involved and has been widely used in various areas such as robotics \cite{meng2008multi} and economy \cite{hodgson2012evolutionary}. There are also many examples of using game theory in intelligent transport systems.
A game theoretic approach for modelling the flow of vehicles in a road with lane change is proposed in \cite{talebpour15}.
%In \cite{banjanovic2016autonomous}, the authors proposed 
A cooperative strategy in non-zero-sum games is applicable to solving conflict situations between two autonomous vehicles in a roundabout \cite{banjanovic2016autonomous}. 
%Li et. al considers 
The decision making based on $k$-level games are studied for two autonomous vehicles at unsignalised intersections \cite{li18} and roundabouts \cite{tian18}.
In our previous work \cite{pruekprasert19}, we proposed a decision making for multiple autonomous vehicles at roundabouts based on Nash equilibria.%, 
%which allows us to consider multiple vehicles at the same time.
%The decision making for two autonomous vehicles \cite{tian18} 
%, unsignalised intersections \cite{li18} and roundabout \cite{banjanovic2016autonomous}. %In our paper, we {\color{red}xxxxx}.

This paper proposes two main contributions. First, we propose a game-theoretic decision making of multiple autonomous vehicles in an unsignalized intersection using Nash equilibria in a perfect information game.
Instead of using the concept of ``aggressiveness'' value proposed in \cite{pruekprasert19}, 
we determine the decision order of the vehicles based on their priorities (e.g., based on the rules of the right of way).
Second, we consider aggressive and unpredictable driving behaviours of malicious vehicles to simulate the real complex traffic environment. Before the widespread adoption of the autonomous vehicles, the human drivers need to share roads with autonomous vehicles. 
%Human's worry and fear of the autonomous vehicle may trigger road rage and result in aggressive driving behaviors. 
For this reason, we classify the drivers using different levels of maliciousness, describing vehicles breaking the law, or having irrational behaviours.
% as {\color{red}xxxxx} to represent the rational level of other vehicles.

%Many scholars do the research on classification of human behaviour, e.g. 

The remainder of this paper is organised as follows. First, we formulate the problem in Section \ref{section:model}. 
The game-theoretic decision making approach is presented in Section \ref{section:decision}. % using game theory. %, which is followed by Section \ref{section:priority}, which
Then, Section \ref{section:priority} introduces the design of formalising priorities. After validating the performance of the proposed method in Section \ref{section:experiments}, Section \ref{section:conclusion} draws the conclusions.
%Finally, some conclusions are drawn in Section \ref{section:conclusion}.

%The prediction of intents of other drivers and the analyses of interaction with other traffic participants 

\section{Problem formulation}\label{section:model}

\subsection{Vehicles at unsignalised intersection}\label{section: scenario}

We consider autonomous vehicles in a four-way unsignalized intersection. We assume that vehicles are driving on the left and that there are at most four vehicles at the intersection, entering from different directions. Their decision making depends on the vehicles that are going to enter or are already engaged in the intersection, but not on those that have already left the intersection. We also assume that every vehicle knows which way every other vehicle is intending to use, as every vehicle declares its intention using their turn signal.

\subsection{Vehicle configurations}\label{section:path}

%========= Must be rewritten more ===============

\begin{figure}[t!]
	\begin{center}	{\includegraphics[width=0.55\linewidth]{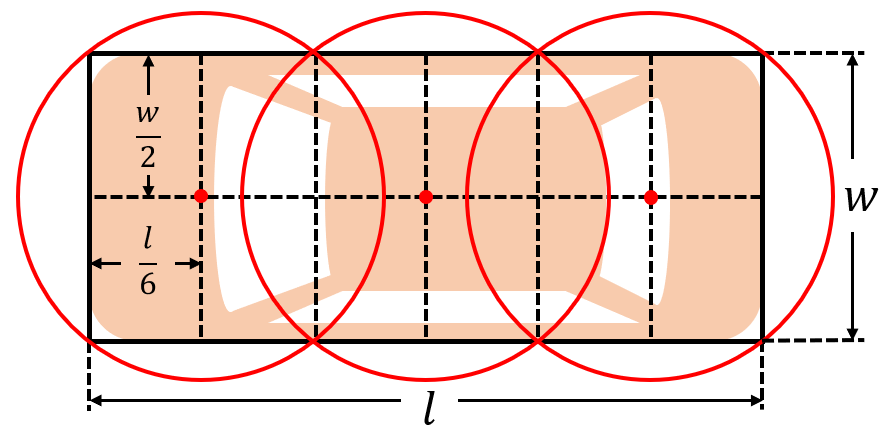}}
		\caption{Road occupancy by over-approximation}
		\label{fig:threecircles}
	\end{center}
\end{figure}

We focus on the high-level decision making of autonomous vehicles following a precomputed path. For this reason, we do not consider the path trackers and the low-level control layer in this work. So then, we over-approximate the road occupancy of each vehicle using three circles (much as \cite{cao18}, see Fig.~\ref{fig:threecircles}) and assume that the vehicles perfectly follow the given
navigation paths.

Much as \cite{pruekprasert19}, we consider the case where the navigation path of each vehicle $i$ is fixed and denoted by $\mathit{path}(i) \in \{$``straight'',``turn right'',``turn left''$\}$. Thus, each vehicle can only control its acceleration along the given navigation path at each time step. More specifically, at each time step $t$, each vehicle $i$ chooses its acceleration $a_i(t)$ in a finite set in order to minimise their overall cost functions.

%This leaves the only control input for each 
%vehicle $i$ at a time step being its acceleration $a_i(t)$ along the given navigation path.
%To simplify the problem, we also assume that each 
%vehicle $i$  chooses an acceleration $a_i(t)$  from a finite set at each time step $t$, so as to minimize their cost functions.

The configuration of the vehicle $i$ at time step $t$ is denoted by:
\begin{equation}
X_i(t)=[x_i(t), y_i(t), v_i(t), a_i(t), \text{sta}_i(t)]^\top
 \label{eq:configuration}
 \end{equation}
where
 $(x_i(t), y_i(t))$ is the position of the vehicle $i$ (where $(0,0)$ is the center of the intersection), $v_i(t)$ is its velocity, $a_i(t)$ is its acceleration, and
  $\text{sta}_i(t) \in \{\text{``entering'' }, \text{``inside'' }, \text{``leaving''}\}$ is its current status (i.e., not yet entered, already inside the intersection, or far enough from the centre of the intersection) respectively.
  
Then, given the control input, that is, the acceleration $a_i(t)$, the configuration at time step $t+1$ is represented by:
\begin{equation}
 X_i(t+1) = \NEXT_i(X_i(t), a_i(t)),
 \label{eq:nextstep}
 \end{equation}
where the function $\NEXT_i$ is computed from the navigation path %is given by the path planner, 
and returns the configuration of the vehicle 
$i$ after one time step, 
 assuming the vehicle has constant acceleration $a_i(t)$ between the 
 time steps $t$ and $t+1$.

\subsection{Rules for the right of way}\label{section:rightofway}

For this paper, we use Japanese rules for deciding priorities of vehicles \cite{npa11}. Rules are slightly different in other countries, especially on the side that have more priority (e.g., left in Japan \cite{npa11}, right in Australia \cite{ntg15}). Each vehicle intending to enter an unsignalized intersection then has to give the right of way to any vehicle (in order of importance):
\begin{itemize}
	\item[A)] already engaged in the intersection,
	\item[B)] on its left-hand side,
	\item[C)] significantly closer to the intersection.
\end{itemize}
A vehicle can proceed and engage in the intersection, as long as it does not cross the way of a vehicle to which it should have given the right of way. In Section \ref{section:priority}, we describe more precisely how we implement these rules in our decision making.

%==================\\
%I decided to make the rules informal here. their precise implementation will be discussed in the next section. The reason is I do not want to talk about randomization here, as it is a bit technical, and is not in the rules\\
%===================\\
%
%===================\\
%===================

\subsection{Levels of maliciousness}\label{section:angelicdemonic}

\begin{figure}[t!]
\begin{center}
{\fontsize{8pt}{8pt}\selectfont
\begin{tabular}{|c|c|c|c|c|}
\hline
type & angelic & inter. & demonic & irrational\\
\hline
initial. & right of way & \multicolumn{2}{c|}{selfish} & \multirow{2}{*}{none}\\
S.\ref{section:initial} & S.\ref{section:rightofway} & \multicolumn{2}{c|}{i.e. highest priority} &\\
\hline
decision & \multicolumn{3}{c|}{\multirow{3}{*}{Nash equilibrium}} & \multirow{3}{*}{random}\\
making & \multicolumn{3}{c|}{} & \\
S.\ref{section:decision} & \multicolumn{3}{c|}{} & \\
\hline
update & right of way & \multirow{3}{*}{fitting} & \multicolumn{2}{c|}{\multirow{3}{*}{none}}\\
priority & + &  & \multicolumn{2}{c|}{}\\
S.\ref{section:updates} & fitting &  & \multicolumn{2}{c|}{}\\
\hline
\end{tabular}}
\end{center}
\caption{Summary of the different level of maliciousness}
\label{figure:malicious}
\end{figure}

In this paper, we consider four kinds of vehicles (see Fig.~\ref{figure:malicious}):
\begin{itemize}
	\item \emph{Angelic}. Such a vehicle tries to follow the rules of Section \ref{section:rightofway} and to avoid any collision.
	\item \emph{Intermediate}. Such a vehicle does not follow the rules from Section \ref{section:rightofway} by initially considering that it has priority. However, it may give priority to other vehicles if the situation requires it.
	\item \emph{Demonic}. Such a vehicle is selfish: it does not respect the rules and believes that it always has the priority. However, it considers other vehicles' behaviours in its decision making.
%	is also selfish, as it will not respect the rules, but may give the right of way to other vehicles if necessary.
	\item \emph{Irrational}. Such a vehicle randomly chooses its acceleration at every time step, independently of the situation.
\end{itemize}

Our goal is to see how our ego vehicle -- considered as angelic -- behaves in presence of different types of vehicles, including some that do not respect the rules, at different degrees of severity.

%=====================\\
%Same as previous section, this should come later. I decided to use angelic/demonic as it is an usual denomination is probabilistic choices. I do not know if this is better. I do not like the name intermediate, but could not find a better name for now\\
%=====================\\
%
%=======================\\
%=======================

\subsection{Structure of the decision making}\label{section:structure}

%Say that there are essentially two components updating the priorities, and computing the control input, that depends on each other. Introduce some notations that will be useful?

In the following two sections, we describe more precisely our decision making for
 autonomous vehicles with different levels of maliciousness. Our method consists of 
 two main 
 components: maintaining priority orders (Section \ref{section:priority}), and computing the control inputs (Section \ref{section:decision}).
  Both components depend on the computations of Nash equilibria (Section \ref{section:game}), similarly to our previous paper \cite{pruekprasert19}.

\section{Control inputs and predictions, as Nash equilibria}\label{section:decision}

Other than irrational vehicles, each vehicle determines its acceleration based on rational decision making.
In this section, we discuss a game-theoretic approach to the problem, similarly to \cite{pruekprasert19}. % of the autonomous vehicles at a roundabout.
%We assume that a vehicle as two objectives: safety and velocity.
%The safety cost is evaluated using the distances between the vehicles, while the velocity cost is computed using the speed of the vehicles. % (see Section \ref{section: cost}).
Using the concept of \emph{priority order} that we will describe more precisely in Section \ref{section:priority}, we formulate the decision making problem as a finite perfect-information game. The players are non-cooperative vehicles, and they try to minimise their cost.
%is modeled as a multi-player game played between the vehicles.
%By using the concept of \emph{aggressiveness},

\subsection{$n$-player game with perfect information}\label{section:game}

We consider \emph{$1$-round sequential games with perfect and complete information} $G = (P,\Gamma,(H_1,\ldots,H_n))$ where:
\begin{itemize}
	\item $P = \{1, \ldots, n\}$ is a finite set of \emph{players},
	\item $\Gamma$ is a finite set of \emph{strategy profiles},
	\item $H_j: \Gamma^n \rightarrow \mathbb{R}$ is the \emph{cost function} for player $j$.
\end{itemize}

In such a game, every player chooses a strategy profile from $\Gamma$ to minimise its cost function. We are particularly interested in \emph{Nash equilibria}, that is, a set of strategies $\gamma_1$, \ldots, $\gamma_n$ for every player that is optimal in the sense that for every player $j$, for every strategy $\gamma_j'$:
\[H_j(\gamma_1, \ldots, \gamma_n) \leq H_j(\gamma_1, \ldots, \gamma_j', \ldots, \gamma_n)\]
When a total order $\prec$ on the set of players is given, it is possible to compute such a Nash equilibrium by \emph{backward induction}: intuitively, the order $\prec$ gives an order with which the players choose their strategy in such a way they ensure optimality.

We invite an interested reader to take a look at usual textbooks, much as \cite{O2009}, for more details on game theory.

\subsection{Cost functions for the decision making in an intersection} \label{section:cost}
%\textcolor{red}{WILL BE CHANGED AFTER CODING!!}
\subsubsection{Cost at each time step}
We first introduce the cost of a vehicle at each time step, which we call the \emph{step-cost}.
The step-cost of the vehicle $i$ is given by
\begin{equation}
%\SCOST_i(\bm{X}, W) = W \cdot \Phi_i(\bm{X}).
\SCOST_i(\bm{X}, \prec) = \phi_i^\mathit{safe}(\bm{X}, \prec)+\phi_i^\mathit{velo}(\bm{X})
\label{eq:stepcost}
\end{equation}
where $\bm{X}$ is a vector of configurations as in (\ref{eq:configuration}), i.e., 
\[\bm{X} = [X_1^\top, \ldots, X_n^\top]^\top\]
and $\prec$ is a total order on the set of vehicles.

%Unless predicted otherwise, $W = [1- a_i ,a_i  ]$ ($a_i$ is the \emph{aggressiveness}).
%where $a_i = 0.9$ (\emph{resp.} $0.1$) if $i$ is  (\emph{resp.} is not) the vehicle with highest priority.
%we use $W = [1- \frac{\PRI(i)}{18} ,\frac{\PRI(i)}{18}  ]$.
%Note that $\PRI(i) \in [0, 15)$, therefore $\frac{\PRI(i)}{18} \in [0 , 0.83\dot{3})$.

%$\Phi_i(\bm{X}) = [\phi_i^\mathit{safe},\phi_i^\mathit{velo}]^\top$.

\subsubsection{Safety}
%Then, we define the safety feature %$\phi_i^\mathit{safe}$ by:
For each pair of vehicles $i$ and $k$, let $d(k,i)$ be the distance between their occupancy, given by the union of three circles (see Section \ref{section:path} and Fig.~\ref{fig:threecircles}).

 With this, we define the \emph{safety feature} by 
 \[\phi_i^\mathit{safe}(\bm{X}, \prec) = \sum_{k \neq i}  \phi_i^k(\bm{X}, \prec),\] 
 where

\begin{equation}
 \phi_i^k(\bm{X}, \prec) =
	\begin{cases}  0 &\mbox{if } \text{sta}_i=\mbox{``leaving''},\\ 
	 \multirow{2}{*}{0} &\mbox{else if the navigation paths}\\
	 &\mbox{of } i \mbox{ and } k \mbox{ do not collide},\\ 
	 0 &\mbox{else if }  d(k,i)\geq D,\\ 
	C_\mathit{d} \cdot (D - d(k,i))^2& \mbox{else if } d(k,i) \leq D_\mathit{danger},\\
	\multirow{2}{*}{0} & \mbox{else if }  i \mbox{ is minimal for } \prec\\
	& \mbox{i.e. } i \mbox{ have highest priority},\\
	C_\mathit{n} \cdot (D - d(k,i))^2 & \mbox{otherwise}
\end{cases}
\label{eq:bfcost}
\end{equation}
%with
%\begin{equation}
%\beta_i^k(\bm{X}) =
%	\begin{cases}
%	0 &\mbox{if } k^* \text{ does not exist},\\ 
%	%\infty &\mbox{if } \theta_f - \theta_i \leq E_{f},\\ 
%	C_\mathit{in,k} \cdot (D - d(k,i))^2 & \mbox{if }\mathit{man}_k =\mathit{enter},\\
% 									& \mbox{and }\mathit{man}_i \neq\mathit{enter},\\
%	C_\mathit{en,k} \cdot (D - d(k,i))^2 & \mbox{if }\mathit{man}_k \neq\mathit{enter},\\
% 									& \mbox{and }\mathit{man}_i=\mathit{enter},\\
%	C  \cdot (D - d(k,i))^2 & \mbox{otherwise}.
% 	 %\frac{1}{  (  \theta_i - \theta_b  -C_f ) ^2} & \mbox{otherwise},
%
%	\end{cases}
%	 \label{eq:beta}
%\end{equation}
%\begin{equation}
%\begin{flalign}
%\nonumber&\phi_i^b(\bm{X},\NEI)=&\\
%&\begin{cases} 0 &\mbox{if }  B = \emptyset,\\ 
	% \infty &\mbox{if }( \theta_i - \theta_b  \leq E_e \text{ and } \mathit{man}_i = \mathit{enter})\\
		%&\mbox{or }( \theta_i - \theta_b  \leq E_b \text{ and } \mathit{man}_i \neq\mathit{enter}),\\
	% C_\mathit{be} \cdot (D - d(f,i))^2&  \mbox{if } \mathit{man}_i = \mathit{enter},\\ 
 %	 C_{b} \cdot (D - d(f,i))^2& \mbox{otherwise},\end{cases}&
%\label{eq:backwardcost}
%\end{flalign}
%\end{equation}
Here, 
$0 < C_\mathit{n} \ll C_\mathit{d}$, $0 < D_\mathit{danger}<D$ are given constants. The intention is that, as long as there is no danger of collision, the vehicle $i$ does not care about safety, but when there is a possibility of collision, the closer the vehicle $k$ is, the more careful the vehicle $i$ is. A special case is when the vehicle $i$ has the highest priority, where $i$ cares less about safety. This case is in order to break symmetric situations where all vehicles would be too conservative.

\subsubsection{Velocity}
Let $v_\mathit{l}$ be the speed limit of the road.
% f(\lvert v_\mathit{limit}- v \rvert)$ where $v(t$ is the velocity of  vehicle $i$ at time-step $t$ and $f$ is a positive increasing convex function.
We define the \emph{velocity feature} by
\begin{equation}
\phi_i^\mathit{velo}(\bm{X}) = \begin{cases}     
C_\mathit{u} \cdot( v_\mathit{l} - v_i)^2 &\mbox{if } v_\mathit{l} \geq v\\
C_\mathit{o}\cdot(v_\mathit{l} - v_i)^2 &\mbox{otherwise}
\end{cases}
\label{eq:velocost}
\end{equation}
Here, we choose $0 < C_\mathit{u} \ll C_\mathit{o}$ so that each vehicle obeys the law.
%where $C_\mathit{in}$, $C_\mathit{en}$ (resp $C_\mathit{o}$) are constant positive coefficients for the cases that $v_{i}$ is under (resp. over) the speed limit. 
%The intention is that $C_o$ is much bigger than $C_\mathit{u}$, because we cannot allow a vehicle to break the law. 
%\textcolor{red}{(Jeremy) I do not see why you need to distinguish ``enter'' from ``inside''. What about when you are in ``exit'' mode?}
 
%Our decision making algorithm uses 
%\subsection{Accumulated cost function for the receding horizon}

 \subsubsection{Accumulated cost function}
We construct the \emph{accumulated cost function} based on the receding horizon control approach~\cite{K2005}, which determines the control inputs of the vehicles based on the predicted future up to a horizon time step $h  < \infty$. 

Given a vector of accelerations $\bm{a} = [a_j(s)]_{j \in N, 0 \leq s \leq h-1}$, define $K_j(\bm{X},\prec,\bm{a}) = \sum\limits_{s = 0}^{h-1}\,\lambda^s\cdot S_j(\widehat{X}(\bm{a},s),\prec)$
where $\lambda$ is the discounted factor and $\widehat{X}(\bm{a},s)$ is defined by induction on $s$:
\begin{itemize}
	\item $\widehat{X}(\bm{a},0) = \bm{X}$,
	\item $\widehat{X}(\bm{a},s+1) = [\NEXT_k(\widehat{X}_k(\bm{a},s),a_k(s)]_{k \in N}$.
\end{itemize}

\subsection{Decision game}\label{section:decisiongame}

In a very similar spirit as \cite{pruekprasert19}, we describe a collection of games whose players are the vehicles at the intersection. Assume that a total order $\prec$ on the set of vehicles and a vector of configurations $\bm{X}$ are given. We define the game $G_{\bm{X},\prec}$ as follows:
\begin{itemize}
	\item The set $P$ of players is the set of vehicles.
	\item The set $\Gamma$ is a finite set of \emph{acceleration patterns}, that is,
	a finite subset of $\mathbb{R}^h$, where $h$ is the time horizon.
	\item For every vehicle $j$, $H_j(\bm{a}) = K_j(X,\prec, \bm{a})$.
	\item The order is given by $\prec$.
\end{itemize}
We can then compute a Nash equilibrium using backward induction as described in Section \ref{section:game}. This produces a collection of 
acceleration patterns that we denote by \[\bm{a}(\bm{X},\prec) = [a_j(s,\bm{X},\prec)]_{j \in N, 0 \leq s \leq h-1}.\]

The intention of $a_j(s,\bm{X},\prec)$ is to be a prediction of the acceleration of the vehicle $j$ after $s$ time steps, starting from the configuration $\bm{X}$, considering $\prec$ as the priority order.

\subsection{Computing control inputs and predicted configurations}\label{section:computation}

Consider a vehicle $i$ that is not irrational. At each time step $t$, the vehicle $i$ observes the precise configuration $\bm{X}(t)$ of every car. It also maintains a \emph{priority order} $\PRIO{i}{t}$, a total order on the set of vehicles. We will see in Section \ref{section:priority} how this order is initialised and updated.

From those data, the vehicle $i$ can consider the game $G_{\bm{X}(t),\PRIO{i}{t}}$, and then compute a Nash equilibrium as previously. We then obtain:
\begin{itemize}
	\item the control input $a_i(t)$ of the vehicle $i$ at time step $t$ given by $a_i(0,\bm{X}(t), \PRIO{i}{t})$,
	\item a prediction $\widehat{a}_{i,j}(t)$ by the vehicle $i$ of the acceleration of the vehicle $j$ at time step $t$.
\end{itemize}

From these predicted accelerations, the vehicle $i$ can compute a prediction of the configuration of the vehicle $j$ at time step $t+1$ by:
\[\widehat{X}_{i,j}(t+1) = \NEXT_j(X_j(t), \widehat{a}_{i,j}(t))\]

\subsection{Resolving deadlocks}\label{section:deadlocks}

The previous subsection described a way of computing the control input by solving a game. This method can however lead to deadlocks, that is, situations where no vehicle takes the decision to go on, and every vehicle waits for others to take the lead. This would typically happen when there is a vehicle in each of the four ways, and that there is no clear priority order that would resolve the situation. Formally, for a vehicle $i$ that is not irrational, this case is to be considered when:
\begin{enumerate}
	\item the velocities of all vehicles are zero, and 
	\item the accelerations of all vehicles are precisely predicted using $\PRIO{i}{t}$.
\end{enumerate}
In the theory of concurrent system, it is known that it is impossible to avoid such a situation using only deterministic choices, and a solution is to introduce some randomness to resolve the situation \cite{lehmann81}.

A vehicle $i$ can unlock the situation in two cases:
\begin{itemize}
	\item either when detecting a deadlock and considering having the highest priority, that is, $i$ being minimal for $\PRIO{i}{t}$,
	\item or when $i$ already detected a deadlock at time $t-1$.
\end{itemize}
In those cases, the vehicle $i$ set its acceleration $a_i(t+1)$ without following a Nash equilibrium. Concretely, in our simulations, this vehicle $i$ can set its acceleration $a_i(t+1)$ to $10m/s^2$ with probability 0.25, which will ultimately break the symmetry of a deadlock situation.

\section{Formalising Priorities}\label{section:priority}

Again, other than irrational vehicles, each vehicle assumes some total order.
In this section, we describe how each of those vehicles $i$ 
initialises and updates a total order $\PRIO{i}{t}$ on the set of vehicles.
This process depends on the type of the vehicle. This is the main 
difference from \cite{pruekprasert19}, where the order $\prec$ was deduced 
from the behaviours of other cars, described by their ``aggressiveness'' values. In this 
paper, the order is obtained from the rules of the right of way, and by the 
character angelic/demonic of the cars.

\subsection{Initialising priorities}\label{section:initial}

\subsubsection{Angelic case}
Let us assume first that the vehicle $i$ is angelic. In this case, the vehicle $i$ will try 
to comply to the rules of the right of way as in Section \ref{section:rightofway}.
For that reason, the vehicle $i$ initialises its priority order $\PRIO{i}{0}$ by randomly choosing a total order that satisfies the following:

\begin{itemize}
\item[A)] If $\text{sta}_j =$ ``inside'' and if $\text{sta}_k \neq$ ``inside'', then $j \prec_{i,0} k$.
\item[B)] Else, if the number of vehicles is less than 4 and $j$ is ``on the left-hand side'' of $k$, $j \prec_{i,0} k$.
%\item Else, the vehicle that goes straight $\prec_i$ left $\prec_i$ right.
\item[C)] Else, let $d_j$ and $d_k$ be the distances of $j$ and $k$ from the centre of the intersection, respectively. If $j$ is significantly closer to the centre of the intersection than $k$, meaning that $d_k - d_j > 2$ meters, then $j \prec_{i,0} k$.
%\item[ Else (same priority $i =_i j$), random the order. 
%If there are more than one vehicle that has the same priority with $i$, say $i =_i j =_i k$, 
%$i$ randomly selects between $i <_i j <_i k$ and $i <_i k <_i j$ and $j <_i i <_i k$ and ...
\end{itemize}

\subsubsection{Selfish cases}
Now, let us assume that the vehicle $i$ is demonic or intermediate. Such a vehicle will consider itself as initially having priority, and will not care about other vehicles. So the vehicle $i$ will initialise its priority order $\PRIO{i}{0}$ by randomly choosing any total order on the set of vehicles such that $i$ is the lowest element $ i \PRIO{i}{0} j$ for all $j\neq i.$

\subsection{Updating the priority predictions}\label{section:updates}
As explained in Section \ref{section:angelicdemonic}, demonic vehicles do not update their priority order, while intermediate and angelic ones do. Thus, if the vehicle $i$ is demonic, we always have
$\PRIO{i}{t+1}   =  \PRIO{i}{t}$ for any time step $t$.
If the vehicle $i$ is angelic or intermediate, it will update its priority only when necessary: when the priority from the rules of the right of way changes and when its predictions are imprecise.

\subsubsection{Update according to the rules of the right of way}
If the vehicle $i$ is angelic, $i$ always updates its priority order whenever the right of way changes. 
For example, when a vehicle leaves the intersection and when a vehicle that does not have priority enters the intersection.
For this case, we update $\PRIO{i}{t+1}$ according to the current right of way in the same way as in Section \ref{section:initial}.
If the rule of the right of way does not change at the current time step $t$, we use $\PRIO{i}{t+1}   =  \PRIO{i}{t}$.

%its priority order is initialised according to the rules of the right of way in Section \ref{section:initial}.
%\textcolor{red}{Assume that the vehicle $i$ is angelic. As we have seen in Section \ref{section:initial}, the rules of the right of rules give some priority order between vehicles. This order is then changed using some fitting to correspond more faithfully to the situation, typically when some vehicles do not respect the law. However, an angelic is meant to respect the right of way, and so it is necessary to reset the priority order so that it matches the priority order given by the rules from Section \ref{section:rightofway} when the situation in the intersection changes. Concretely, this happens when a vehicle leaves the intersection or a vehicle that did not have priority enters the intersection.}

\subsubsection{Update due to imprecise predictions}
We consider this type of update for intermediate vehicles and angelic vehicles that do not use the update according to the rules of the right of way at the current time step $t+1$. Suppose that $i$ is one of those vehicles.
As in Section \ref{section:computation}, given the priority order $\prec_{i,t}$ and the observed configuration $\bm{X}(t)$ at the previous time step $t$, the vehicle $i$ can predict the configurations $\widehat{X}_{i,j}(t+1)$ of any other vehicle $j$. 
In the case where these predictions are different from the observed configurations $X_j(t+1)$, the vehicle $i$ updates its priority order to fit more closely these new observations. To this end, the vehicle $i$ computes the Nash equilibrium of the decision game $G_{\bm{X}(t), \prec}$ for any total order $\prec$, and choose $\prec'$ as the total order such that:
\begin{itemize}
	\item The difference between predicted and observed accelerations is minimal:
	\[\prec' \in \arg\min_{\prec'}\sum\limits_{j \in N} \lvert a_j(0,\bm{X}_j(t), \prec') - a_j(t+1)\rvert.\]
%	
%	The predicted configuration
%	\[\NEXT_j(X_j(t), a_j(0,\bm{X}_j(t), \prec'))\]
%	is the closest to the observation $X_j(t+1)$,
	\item If several $\prec'$ satisfy the first condition, choose the one that gives the minimal acceleration $a_i(0,\bm{X}(t), \prec')$.
\end{itemize}
The intention of the second condition is the following: the reason why vehicle $i$ needs to change its priority order is because some vehicles are going against its believes (either the rules of the right of way for angelic, or selfishness for intermediate). In any case, the vehicle has to be particularly careful in the case it is dealing with a demonic or an irrational vehicle. We also define $\PRIO{i}{t+1}$ as follows:
\begin{itemize}
	\item If $a_i(0,\bm{X}(t), \prec') \leq a_i(0,\bm{X}(t), \PRIO{i}{t})$, $\PRIO{i}{t+1} \,=\, \prec'$.
	\item Otherwise, $\PRIO{i}{t+1} \,=\, \prec'$ with probability $0.25$, and $\PRIO{i}{t+1} ~=~ \PRIO{i}{t}$ with probability $0.75$.
\end{itemize}
Again the intention of the second case is to be careful in a situation where the vehicle would benefit by being more aggressive, while in a situation that contradicts its believes.

\section{Experimental results}\label{section:experiments}
%In this section, we presents the numerical simulation results to verify the effectiveness of the proposed approach.
To verify the effectiveness of our approach, we perform numerical simulations using Matlab 2018a and 2018b.

\subsection{Experiment scenario}

\begin{figure}
\centering    
  \includegraphics[ height=4cm]{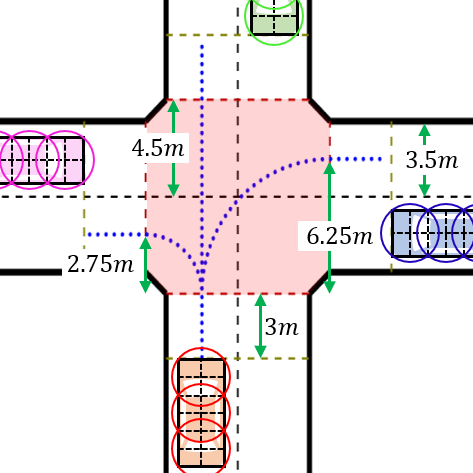}
\caption{The unsignalized intersection used for the simulations, the three navigation paths, and the initial positions of each vehicle.\label{fig:intersection}} 
\end{figure}

We consider four vehicles approaching an unsignalized intersection as shown in Fig.~\ref{fig:intersection}: each vehicle approaching from each entrance.
We assume that all vehicles have rectangle shapes, and over-approximate the road occupancy of each vehicle using three circles shown in Fig.~\ref{fig:threecircles}. 
We consider that two vehicles collide if their road occupancy areas intersect. %each other. %, that is, at least one of their circles intersects any of the circles of the other vehicle.

%``entering''$ if the vehicle $i$ is approaching the intersection but its part intersect the the intersection.

%\in \{\text{``entering'' }, \text{``inside'' }, \text{``leaving''}\}$

The structure of the intersection and the three navigation paths are also illustrated in Fig.~\ref{fig:intersection}.
%The initial velocity of all vehicles is zero. 
The navigation path type of each vehicle is randomised.
The length and width of each vehicle are randomised within the range $(3.5m, 5.5m)$ and $(1.5m,2.1m)$, respectively.
We consider that the navigation paths of given two vehicles do not collide only when 
the two vehicles approach the intersection from the opposite directions, and each vehicle either turns left or goes straight.

For each vehicle $i$, $\text{sta}_i(t)$ is given as follows:
\begin{enumerate}
\item $\text{sta}_i(t)= $ ``entering'' if the vehicle $i$ is approaching the intersection but none of its part has yet entered the intersection, i.e., the rectangle that represents the vehicle has not yet intersected the red area in Fig.~\ref{fig:intersection}.
\item $\text{sta}_i(t)= $ ``leaving'' if more than half of the vehicle is already outside the intersection.
\item $\text{sta}_i(t)= $ ``inside'' otherwise.
\end{enumerate}

The decision making is performed every $0.1s$. 
The discounted factor for the accumulated cost function is $\lambda=0.8$.
The speed limit of the road is $v_l =16.7 m/s$. %\cite{dmvca}.
The constant parameters are as follows:
$C_{n}=20$, $C_{d}=10^{300}$, 
$D = 25\,m$, $D_\mathit{danger} = 0.5\,m$,
$C_u = 1$, and $C_o = 1000$.

We use the following patterns of acceleration/deceleration sequences
with a time horizon $h = 3$. All the accelerations in this section are in $m/s^2$.
 % \textcolor{red}{check}: 
		\begin{itemize}
			\item $[-50, -50, -50]$ for a deceleration,
			\item $[0,0,0]$ for no acceleration,
			\item $[10,0,0]$ for a small acceleration,
			\item $[20, 0, 0]$ for a strong acceleration.
		\end{itemize}

The accelerations of a irrational vehicle are randomly selected from the set $\{-50,0,10,20\}$. 

\subsection{Results and analysis}
\begin{table}
\begin{center}
\caption{Simulation Results}
\begin{threeparttable}
\renewcommand{\arraystretch}{1.8}
{\fontsize{6.5pt}{6.5pt}\selectfont
\newcommand{\tabincell}[2]{\begin{tabular}{@{}#1@{}}#2\end{tabular}}
\begin{tabular}{r|ccc}
\hline
\textbf{Case} & \textbf{Collision rate(\%)} & \textbf{Congestion rate(\%)} & \textbf{Avg. Total time steps}\\
\hline
1 & 0 & 0 & 56.87 (5.687s)\\
\hline
2 & 0 & 0.2 & 53.98 (5.398s)\\
\hline
3 & 0 & 0 & 59.09 (5.909s)\\
\hline
4 & 0.4 & 4.0 & 91.88 (9.988s)\\
\hline
1' & 0 & 0.5 & 55.43 (5.543s)\\
\hline
2' & 0 & 1.4 & 50.58 (5.058s)\\
\hline
3' & 0 & 9.4 & 55.81 (5.581s)\\
\hline
4' & 1.1 & 14.3 & 75.82 (7.582s)\\
\hline
\end{tabular}}
%\vspace{0.05cm}
%\begin{tablenotes}
%    \tiny{\item[1] Average total time steps of 998 simulations. Two simulations took more than 500 time steps (timeout).}
%  \end{tablenotes}
\end{threeparttable}
\end{center}
%{\tiny{$^1$ Average total time steps of 998 simulations. The other two simulations take more than 500 time steps (timeout).}}\\
\label{Table:ExperimentalResult}
\end{table}

%480 510
%

We perform 2000 simulations for all those cases.
\begin{enumerate}
\item[1] Four angelic vehicles. 
%\item[2] Two angelic vehicles and two intermediate vehicles. 
\item[2] Three angelic vehicles and one demonic vehicle. 
%\item[4] Three angelic vehicles and one irrational vehicle. 
\item[3] Four intermediate vehicles.
\item[4] Three intermediate vehicles and one irrational vehicle. 
\end{enumerate}
1000 simulations are done with initial velocity $v_i(0) = 0$ (cases 1-4), and 1000 simulations with random initial velocity depending on the type of the vehicle (cases 1'-4'):
\begin{itemize}
	\item if the vehicle is demonic or irrational, we randomly choose in the interval $[0,16.7]$m/s,
	\item if the vehicle is angelic or intermediate, we randomly choose in the interval $[0,6]$m/s.
\end{itemize}

Observe that cases 3 and 4 contain no angelic vehicles, even though we assume that the ego vehicle should be angelic. 
This is to see how robust the fitting method of Section \ref{section:updates} is for intermediate vehicles.
%The reason is that we wanted to see how robust the fitting method of Section \ref{section:updates} is for intermediate vehicles.
 
As presented in Table  \ref{Table:ExperimentalResult}, we evaluate the simulations by considering the following criteria.
The first column presents the collision percentage.
The second column presents the percentage of the simulations that have congestions, which are situations such that: 1) there are at least two vehicles, $i$ and $j$, such that  
$\text{sta}_i(t)= \text{sta}_j(t)= \mbox{``inside''}$ and 2) the navigation paths of both vehicles may collide.
The third column presents the average total time that the vehicles spend running along their navigation paths.

For cases 1, 2, and 3, all vehicles successfully leave the intersection without any collisions.
%However, the congestion rate of the case 3, where all vehicles do not respect the rules for the right of way, is significantly higher than cases 1 and 2.
Some collisions are detected for the case 4, due to the unpredictable behaviour of the irrational vehicle.
In the cases where the initial velocity is possibly non zero, we detect more congestions, due to the fact that vehicles have less time to make their decision before entering the intersection. However, the total time is shorter because each vehicle starts with some initial velocity.

The results also show that the vehicles may spend less time in the intersection if some of them are selfish.
The case 2 is particularly interesting because it has fewer time steps than case 1.
One possible explanation for this case is that the selfish vehicle makes the decision making easier: it will force itself through the intersection, so that angelic vehicles have no choice but to let it go first. After this vehicle leaves the intersection, the decision making with three vehicles is much easier.

 \begin{figure}[t]
    \centering
      \includegraphics[scale=0.14]{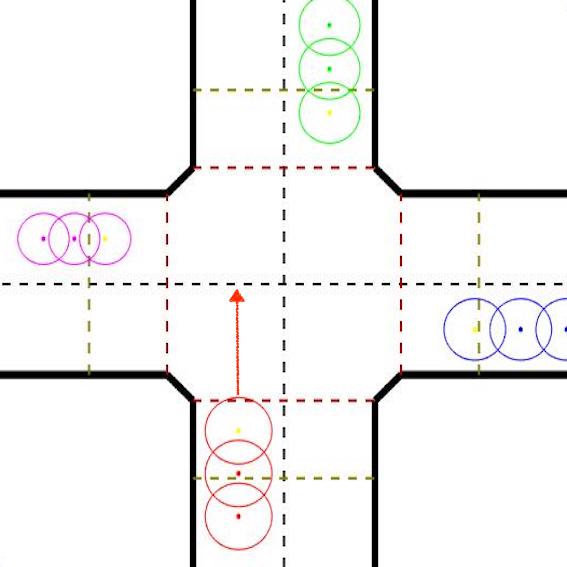}
      \includegraphics[scale=0.14]{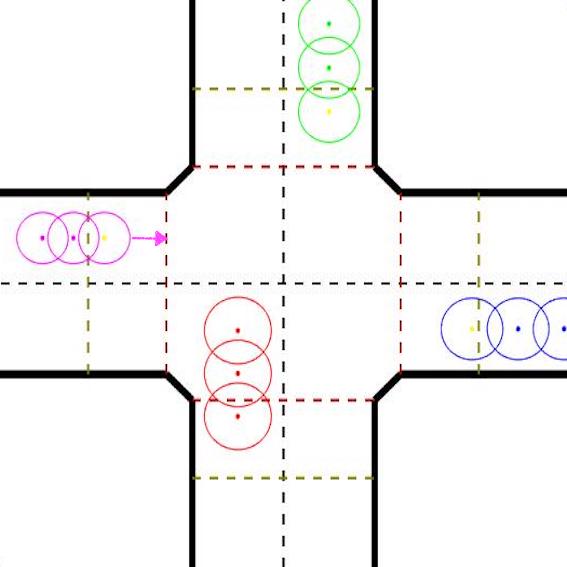}
      \includegraphics[scale=0.14]{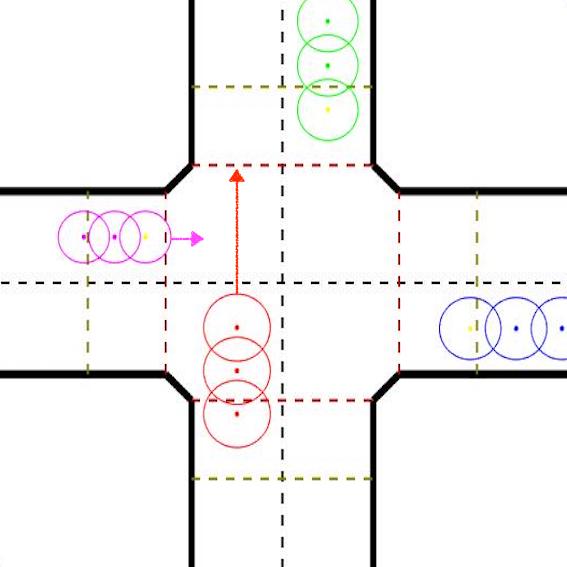}
     {\scriptsize   (a) time step = 9  \hspace{1cm} (b) time step = 16  \hspace{1cm} (c) time step = 20}
    \caption{A dangerous situation%(a) At the time step 4, the pink (angelic) vehicle is inside the intersection.
    %(b) At time step 5, the red (irrational) vehicle move forward. (c)
    \label{fig:example}
    }
  \end{figure}

In Fig.~\ref{fig:example}, we present a dangerous situation of the case 4'.
The red vehicle is irrational, while the other three vehicles are intermediate.
In Fig.~\ref{fig:example}(a), the red vehicle is entering the intersection, while the other three are stopped.
However, in Fig.~\ref{fig:example}(b), the red vehicle suddenly stops, so the pink vehicle thinks it can enter the intersection.
Finally in Fig.~\ref{fig:example}(c), the red vehicle starts again and the pink vehicle is forced to update its priority order as in Section \ref{section:updates} and gives the way to the red vehicle. %After the red vehicle leaves the intersection, the pink vehicle continues running along its navigation path.
%Then, the remaining two vehicles continue moving forward according to the rules for the right of way as in Section \ref{section:rightofway}, i.e., the blue vehicle goes before the green one. 
This scenario demonstrates that our decision making is robust enough to avoid such potential accidents.

\section{Conclusion}
\label{section:conclusion}

In this paper, we propose a decision making for autonomous vehicles in a unsignalized intersection in presence of selfish and irrational vehicles. The decision is made by computing Nash equilibria of sequential games played by the vehicles. The decision orders of those games are based on the believes of which vehicles have priority: angelic vehicles define their priorities based on the rules of the right of way, while malicious ones regard themselves as having priority or do not pay any attention to priorities. 
We consider several scenarios where vehicles respecting the law (angelic) share the road with selfish vehicles and irrational ones, which randomly choose their accelerations. We illustrate those scenarios in our numerical simulations, demonstrating its feasibility and the robustness of our decision making in presence of malicious vehicles. As a future work, systematic experiments based on more sophisticated and practical traffic simulators will be conducted to further evaluate our model, and more detailed techniques will be adopted to eliminate the potential hazards. %Moreover, algorithm with high efficiency (e.g. Monte carlo tree search) will be designed  to improve the efficiency to search for the Nash Equilibrium. In addition, machine learning techniques (e.g. transfer learning) will be considered to adapt crossing situations with various configuration.

%we would like to confront our decision making with human drivers.

\bibliographystyle{IEEEtran}

\end{document}